\documentclass[reprint,twocolumn,aps,prB,amsmath,amssymb,floatfix,superscriptaddress,longbibliography]{revtex4-1}
\usepackage{graphicx}
\usepackage{epsfig}
\usepackage{bm}
\usepackage{dcolumn}
\usepackage{color}
\usepackage{physics}
\usepackage{float}
\usepackage[colorlinks,urlcolor=blue,citecolor=blue,linkcolor=magenta]{hyperref}
\makeatletter
\newcommand*{\rom}[1]{\expandafter\@slowromancap\romannumeral #1@}
\makeatother
\usepackage{tikz}
\usepackage{subfigure}

\usepackage{diagbox}
\usepackage{booktabs}
\usepackage{multirow}
\usepackage{makecell}
\usepackage{amsfonts}
\usepackage{amssymb}
\usepackage{graphicx}
\usepackage{dcolumn}
\usepackage{bm}
\usepackage{amsmath}
\usepackage{graphicx}
\usepackage{subfigure}
\usepackage{float}
\usepackage{color}

\begin{document}

\preprint{APS/123-QED}
\preprint{This line only printed with preprint option}

\title{Hidden Unbounded Potential and Re-Entrant Multifractalization in a Generalized Su-Schrieffer-Heeger Model}

\author{Yun-Yan Chen}
\affiliation {Key Laboratory of Atomic and Subatomic Structure and Quantum Control (Ministry of Education), Guangdong Basic Research Center of Excellence for Structure and Fundamental Interactions of Matter, School of Physics, South China Normal University, Guangzhou 510006, China}

\author{Jia-Ming Zhang}
\affiliation {Key Laboratory of Atomic and Subatomic Structure and Quantum Control (Ministry of Education), Guangdong Basic Research Center of Excellence for Structure and Fundamental Interactions of Matter, School of Physics, South China Normal University, Guangzhou 510006, China}
\affiliation {Guangdong Provincial Key Laboratory of Quantum Engineering and Quantum Materials, Guangdong-Hong Kong Joint Laboratory of Quantum Matter, Frontier Research Institute for Physics, South China Normal University, Guangzhou 510006, China}

\author{Zhi Li}
\email[Corresponding author: ]{lizphys@m.scnu.edu.cn}
\affiliation {Key Laboratory of Atomic and Subatomic Structure and Quantum Control (Ministry of Education), Guangdong Basic Research Center of Excellence for Structure and Fundamental Interactions of Matter, School of Physics, South China Normal University, Guangzhou 510006, China}
\affiliation {Guangdong Provincial Key Laboratory of Quantum Engineering and Quantum Materials, Guangdong-Hong Kong Joint Laboratory of Quantum Matter, Frontier Research Institute for Physics, South China Normal University, Guangzhou 510006, China}

\date{\today}

\begin{abstract}

We study the multifractal criticality in a generalized Su-Schrieffer-Heeger model. The results show that the system supports not only critical phases but also re-entrance multifractalization (REM). By mapping the hopping term to an effective potential, we analytically prove that although the model has no explicit unbounded potential, a hidden unbounded potential is actually present—this is the key mechanism driving the emergence of multifractal critical phases. Moreover, one can get a condition where the competition between the explicit and hidden unbounded potentials is exactly balanced. Under this condition, the multifractal critical phase vanish, and the system returns to the extended phase. Based on this mechanism, we achieve both demultifractalization and re-entrant multifractalization. Finally, we double check the theoretical predictions through wave packet dynamics, and the numerical results are consistent with our theoretical analysis. This work broadens our understanding of how unbounded potentials induce multifractal critical phases, providing a theoretical basis for designing new systems with multifractal critical phases.
\end{abstract}

\maketitle

\section{Introduction}

Anderson's localization theory points out that random disorder is the cause behind the dynamical freezing of electron transport~\cite{PWAnderson1958}. This theory uncovers the microscopic mechanism of the metal-insulator transition, promoting the development of the semiconductor industry over the past few decades up to the present. Through scaling theory~\cite{EAbrahams1979, PALee1985, BHetenyi2021}, we know that in low-dimensional (1D and 2D) cases, even tiny random disorder can lead to the localization of all electronic states, which macroscopically manifests as zero conductivity. However, in high-dimensional (e.g., 3D) systems, electron eigenstates exhibit a critical point for phase transition: when the disorder is weak, the system is in the delocalized phase; when the disorder is strong, the system enters the localized phase. Besides, high-dimensional systems also feature energy-dependent eigenstates, from which the mobility edge emerges~\cite{FEvers2008, ALagendijk2009, NMott1987}.

Unlike random disorder, quasiperiodic disorder lies between random disorder and order. Systems with quasiperiodic disorder can induce mobility edges even in low-dimensional cases. Moreover, from a mathematical perspective, the simple function structure of quasiperiodic systems gives them the advantage of being exactly solvable~\cite{DJThouless1988, MKohmoto1983, MKohmoto2008, XCai2013, GRoati2008, YLahini2009, DTanese2014, HPLuschen2018, FAAn2018,FAAn2021, YWang2022a, HLi2023, TYLi2024}. The Aubry‑Andr\'e (AA) model, as one of the most typical one-dimensional quasiperiodic systems, has garnered considerable interest owing to its self-duality and exact solvability.~\cite{SAubry1980, PGHarper1955, MGonalves2022}. Over recent years, many studies based on AA model has uncovered a wealth of intriguing phenomena such as the critical phase, multifractal enriched mobility edge and mobility ring~\cite{Avila2017AC, MGon2023, HLi2023, YHatsugai1990, JHHan1994, YTakada2004, FLiu2015, JWang2016, SZLi2025, SZLi2024b} by means of modifying quasiperiodic potentials~\cite{SDSarma1988, SDSarma1990, ZLu2022, SLZhu2013, YWang2020a, SGaneshan2015, HYao2019, XLi2020, TLiu2022, XPLi2016, XLi2017, BFZhu2023, EWLiang2023, SJitomirskaya2017, YCZhang2022, ZBWang2022}, introducing hopping terms~\cite{JBiddle2010, JBiddle2011, XDeng2019, XXia2022, MGon2023, XCZhou2023a}, or extending to non-Hermitian frameworks~\cite{YJZhao2025, SZLi2024a, SZLi2024b, GJLiu2024, SZLi2024c, SLJiang2023, DWZhang2020a, DWZhang2020b, HJiang2019, JLDong2025, QLin2022, TLi2022}. These advances have broadened the horizon of quantum phase transitions in disordered systems.

Multifractal critical states (MCS) have attracted increasing attention due to their potential correlation with the enhancement of superconducting transition temperature~\cite{JMayoh2015,MVFeigelman2007,MVFeigelman2010,ZFan2021,XZhang}. Typically, critical states only emerge at the critical point of the extended-localized phase transition.  However, recent studies have revealed that in specific models, the critical point can expand into a region, which is known as the critical phase, and is also termed a multifractal phase due to the fractal characteristics of the wave functions.~\cite{Avila2017AC, MGon2023, HLi2023, YHatsugai1990, JHHan1994, YTakada2004, FLiu2015, JWang2016, SZLi2025}. From a dynamical perspective, the localized state wave function remains frozen over time, while the extended state wave function undergoes ballistic transport over time and eventually become uniformly spread over the entire space. The multifractal state lies between the two, characterized by anomalous power-law diffusion~\cite{MHopjan2025, SAbe1987, JXZhong2001, ZJZhang2012, JLDong2024}. From the perspective of the energy spectrum structure, the extended, localized, and multifractal phases correspond to the absolutely continuous (AC) spectrum, the pure point (PP) spectrum, and the singular continuous (SC) spectrum, respectively~\cite{Avila2017AC}. Given this unique nature, two methods have been developed for constructing multifractal phases: one is to introduce unbounded potential modulation satisfying the Simon-Spencer theorem in the potential term~\cite{BSimon1989, SLonghi2023}; the other is to introduce generalized non-commensurate zeros in the hopping term~\cite{Avila2017AC, XCZhou2023a, XCZhou2025}. Both methods essentially cut the AC spectrum into SC spectra~\cite{Avila2017AC, XCZhou2023a, BSimon1989}, thereby achieving the multifractal phase. Then, it is natural to wonder that the generality of this mechanism, i.e., whether it can still take effect if these elements are not explicitly included in the Hamiltonian?

On the other hand, the traditional view holds that when a system undergoes the localization transition, the localized state will remain stable even if disorder continues to increase. However, recent research finds that in the generalized AA model, there is a new non-monotonic phase transition phenomenon that depends on the disorder intensity, i.e., re-entrant localization (REL)~\cite{VGoblot2020, LJ2021, AStrkalj2021, APadhan2022, SRoy2021, SZLi2023}. In concrete terms, as the disorder intensity continues to increase, the REL phase transition is manifested as the system repeatedly entering and jumping out of the localized phases. This goes beyond the previous understanding of Anderson localization. Inspired by recent research on multifractal phase and re-entrant localization, another natural question arises: Is it possible to achieve re-entrant multifractalization (REM) in systems with multifractal phase? This paper aims to address these two issues.

The rest of this manuscript is organized as follows:
In Section~\ref{Sec2}, we introduce our model and define the key observable quantities. Section~\ref{Sec3} investigates the emergence mechanism of multifractal phase in the model, presenting both numerical and analytical results. The corresponding de‑multifractalization mechanism is provided in Section~\ref{Sec4}. Utilizing these findings, we construct a re‑entrant multifractalization model in Section~\ref{Sec5}. Finally, we summarize our conclusions in Section~\ref{Sec6}.

\section{Model and key observables}\label{Sec2}

\begin{figure}[htbp]
\centering
\includegraphics[width=8.5cm]{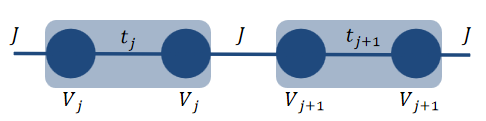}
\caption{Schematic diagram of the model in Eq.~\eqref{E1}.}
\label{fig-1}
\end{figure}

Let's start at a generalized Su-Schrieffer-Heeger model with quasiperiodic modulation both on potential and hopping terms. The corresponding Hamiltonian reads,
\begin{equation}\label{E1}
H=\sum_{j=1}^{L/2}(t_ja_{j}b_{j}^\dag+Jb_{j}a_{j+1}^\dag+H.c.)+\sum_{j=1}^{L/2}V_j(b_jb_j^\dag+a_ja_j^\dag),
\end{equation}
where
\begin{equation}\label{E2}
V_j=\frac{\lambda}{\cos(2\pi\alpha j+\theta)}~~\&~~t_j=\frac{t}{\cos(2\pi\alpha j+\theta)}.
\end{equation}
Here, $a_{j}^{\dagger}$ and $b_{j}^{\dagger}$ ($a_{j}$ and $b_{j}$) are creation (annihilation) operators on $a$ and $b$ sublattices of the $j$-th site, respectively. $\lambda$ is the strength of the quasiperiodic potential and $L$ is the system size. $t_j$ and $J$ correspond to the strength of intra- and extra-hopping of unit cells~(see Fig.~\ref{fig-1}). $\alpha$ and $\theta$ are quasiperiodic parameter and global phase. Without loss of generality, we set $\theta=0$ and $\alpha=\frac{\sqrt{5}-1}{2}$ in the following calculations. It should be noted that this model differs from the paradigmatic quasiperiodic AA model in two respects: (i) not only the potential but also hopping is qiasiperiodic, and (ii) the cosines are in the denominator, to make a model with unbounded potential and hopping.

Next, we define the time-independent observables employed in this work. Here we present the expressions necessary for the numerical calculations; a detailed description is provided in Appendix~\ref{AppA}.

Since our subsequent discussion focuses on the localization properties, the inverse participation ratio (IPR) and normalized participation ratios (NPR) are frequently employed~\cite{GADominguez-Castro2019, APadhan2022, XLi2017, SRoy2021, XLi2020}, denoted as $\xi_2$ and $\zeta_2$, respectively, are defined by
\begin{equation}
\xi_2(\beta)=\frac{\sum_{j=1}^L\vert\psi_{j}(\beta)\vert^4}{(\sum_{j=1}^L\vert\psi_{j}(\beta)\vert^2)^2},
\end{equation}
and
\begin{equation}
\zeta_2(\beta)=\frac{1}{L\xi_2(\beta)}.
\end{equation}
where $\vert\psi_j(\beta)\rangle=\sum_j\psi_j(\beta)\vert j\rangle$ for the $\beta$th eigenstate.

Fractal dimension~\cite{HYao2019, XDeng2019, YWang2020a, YWang2022b, SZLi2024c, AJagannathan2021}, which serves as a core physical quantity reflecting the localization properties, can be derived from $\xi_2$ and is defined as
\begin{equation}\label{E5}
\Gamma_2(\beta)=-\lim_{L\rightarrow\infty}\frac{\ln\xi_2(\beta)}{\ln L},
\end{equation}
where $\Gamma\to0$ ($\Gamma\to1$) corresponds to a localized (extended) state, while $0<\Gamma<1$ corresponds to a multifractal state. Note that, the fractal dimension $\Gamma$ equals to $0$ ($1$) for the localized (extended) state under the condition of thermodynamic limit $(L=\infty)$, while the fractal dimension $\Gamma$ of the multifractal state in the thermodynamic limit remain between $0$ and $1$. The thermodynamic limit results can be obtained by conducting a extrapolation fitting~\cite{YWang2022b}. To get corresponding properties in the localized and extended phase region, we define the average fractal dimension, i.e.,
\begin{equation}\label{E6}
\overline\Gamma_2=\frac{1}{N[R]}\sum_{R}\Gamma_2(\beta), 
\end{equation}
where $R=$Loc., Ext., or MF. represents the eigenstates in localized, extended, or multifractal  regions. $N[R]$ denotes the total number of eigenstates in region $R$.

Using $\xi_2$ and $\zeta_2$, we can define another physical quantity $\eta$, which read~\cite{APadhan2022, XLi2017, SRoy2021, XLi2020}
\begin{equation}\label{E7}
\eta=\ln(\frac{1}{N[\beta/L\in\mathcal{R}]^2}\sum_{\beta/L\in\mathcal{R}}\xi_2(\beta)\times\sum_{\beta/L\in\mathcal{R}}\zeta_2(\beta)),
\end{equation}
where $\mathcal{R}$ represents the range from which eigenstates are selected. In subsequent calculations, we select $\mathcal{R}=[1-\alpha,\alpha]$. Consistent with the preceding text, $N[\beta/L\in\mathcal{R}]$ denotes the total number of eigenstates in region $\beta/L\in\mathcal{R}$.

For a purely localized or extended phase, $\eta\approx\ln(1/L)$. In contrast, for a mixed phase consisting of both localized and extended states, one has $\eta\gg\ln(1/L)$ in the large-$L$ limit~\cite{APadhan2022, XLi2017, SRoy2021, XLi2020}. Moreover, in a multifractal phase, since the fractal dimensions of individual eigenstates are generally not equal to one another, the system exhibits features resembling those of a mixed phase. As a consequence, the value of $\eta$ computed for a multifractal phase becomes significantly larger than that for a localized or extended phase, thereby allowing the multifractal phase to be distinguished from other phase regions.

Finally, the three different phases can also be identified via the characteristics of the spectrum. To this end, the even-odd(odd-even) level spacings are defined as~\cite{APadhan2022, YZhang2022, MSarkar2021, XDeng2019, RQi2023}
\begin{equation}
\delta_n^{e-o,o-e}=E_{2n}-E_{2n-1},E_{2n+1}-E_{2n}.
\end{equation}
Here, $n=1,2,3,\dots,n_{max}$, denotes the index of the even-odd (odd-even) level spacings arranged in ascending order. In the extended phase, the spectrum typically exhibit doubly degeneracy feature~\cite{SAubry1980}, thereby directly implying that $\delta_n^{e-o}\gg\delta_n^{o-e}$ or $\delta_n^{o-e}\gg\delta_n^{e-o}$. For the localized phase, on the other hand, the doubly degeneracy is lifted, resulting in $\delta_n^{e-o}\approx\delta_n^{o-e}$. To differentiate the phases even in the thermodynamic limit, we take the logarithm of the two sets of level spacings, denoted as $\ln\delta_n^{e-o}$ and $\ln\delta_n^{o-e}$. For the extended phase, a noticeable gap emerges between them; for the localized phase, the gap disappears; and for the multifractal phase, due to its critical behavior intermediate between the extended and localized phases, the two sets exhibit a scattered distribution distinctly different from the previous two characteristics~\cite{XDeng2019}.

\section{Emergence of multifractal phase}\label{Sec3}

\begin{figure}[htbp]
\centering
\includegraphics[width=8.5cm]{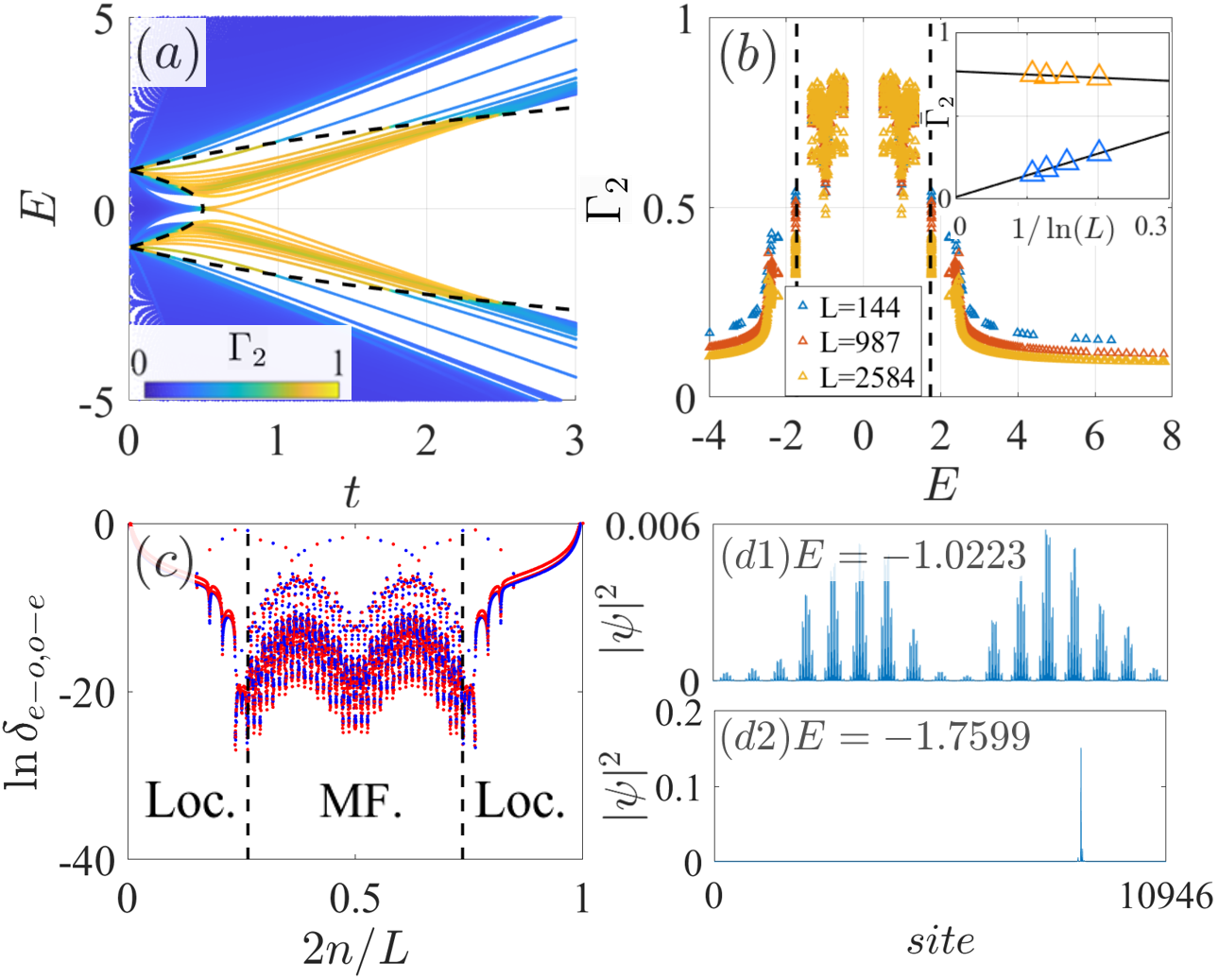}
\caption{(a) Phase diagrams of fractal dimensions in $t$-$E$ plane with $\lambda=0$. It can be seen that there exist two regions (multifractal or localized) where $\Gamma_2$ exhibits distinctly different behaviors, and these two regions are precisely separated by the mobility edge (black dashed line) described by Eq.~\eqref{E18}. (b) The $t=1$ cross profile of (a) for different system size.  The fractal dimension of the localized region converges to 0 with increasing system size, while the multifractal region displays no apparent regularity. The inset exhibits the scaling behavior of multifractal (brown triangles) and localized (blue triangles) regions by extrapolation fitting of the mean fractal dimension of $L=$ 144, 610, 2584 and 10946. (c) The level spacings $\ln\delta^{e-o}$ (blue) and $\ln\delta^{o-e}$ (red) under the condition of $\lambda=0$ and $t=1$. Only the scattered distribution characteristic of the multifractal phase and distinct gap characteristic of the localized phase are observed}. (d) Density distribution of eigenstate in multifractal (d1) and localized (d2) region of (a). Throughout, we set $J=1$ as the energy unit. $L=2584$ for (a), while $L=10946$ for (c),(d1) and (d2).
\label{fig-2}
\end{figure}

Now, Let's first discuss the extreme case of $\lambda=0$. Notably, under this condition, the system contains neither unbounded potentials nor incommensurate zeros hopping, which are two main ways to implement multifractal phase. However, the results still indicate the emergence of a multifractal phase~(see Fig.~\ref{fig-2}(a)). In concrete terms, the numerical results of the fractal dimension exhibits both localized and multifractal behaviors~\cite{HYao2019, XDeng2019, YWang2020a, YWang2022b, SZLi2024c, AJagannathan2021}, separated by a mobility edge (black dashed line) described by Eq.~\eqref{E18}. To verify the robustness of these findings across varying system sizes, we conducted a finite-size analysis~\cite{YWang2022b}. The analysis confirms that, the fractal dimension of the multifractal region converges to a value between $0$ and $1$ with an increasing system size $L$, indicating the emergence of a stable, size‑independent multifractal phase (see Fig.~\ref{fig-2}(b)). Furthermore, the even‑odd (odd‑even) level spacings (Fig.~\ref{fig-2} (c)) in the multifractal region exhibit neither a distinct gap nor complete closure, but instead show a clear scattered distribution, providing an additional evidence for the emergence of the multifractal phase~\cite{APadhan2022, YZhang2022, MSarkar2021, XDeng2019, RQi2023}. Finally, we display the spatial profiles of the wave functions in critical (Fig.~\ref{fig-2} (d1)) and localized (Fig.~\ref{fig-2} (d2)) phases, respectively. This once again confirms the conclusion of multifractal phase emergence.

For the model Eq.~\eqref{E1}, we can provide a theoretically exact solution. From the Hamiltonian Eq.~\eqref{E1}, one can obtain the corresponding eigenequation set, i.e., 
\begin{equation}
\begin{matrix}
E\psi_{a,j}=J\psi_{b,j-1}+V_j\psi_{a,j}+t_j\psi_{b,j},\\
\\
E\psi_{b,j}=t_j\psi_{a,j}+V_j\psi_{b,j}+J\psi_{a,j+1}.
\end{matrix}
\end{equation}
The corresponding transition matrix reads
\begin{equation}
\begin{split}
T_{j}(\theta)&=
\left(
\begin{array}{cc}
\frac{E-V_j}{J}&-\frac{t_j}{J}\\
1&0
\end{array}
\right)
\left(
\begin{array}{cc}
\frac{E-V_j}{t_j}&-\frac{J}{t_j}\\
1&0
\end{array}
\right)\\
&=
\left(
\begin{array}{cc}
\frac{(E-V_j)^2}{Jt_j}-\frac{t_j}{J}&-\frac{E-V_j}{t_j}\\
\frac{E-V_j}{t_j}&-\frac{J}{t_j}
\end{array}
\right).
\end{split}
\end{equation}
Substituting Eq.~\eqref{E2} with $\lambda=0$ into the above expressions yields
\begin{equation}
T_{j}(\theta)=\frac{1}{\Theta_j}
\left(
\begin{array}{cc}
\frac{E^2}{Jt}\Theta_j^2-\frac{t}{J}&-\frac{E}{t}\Theta_j^2\\
\frac{E}{t}\Theta_j^2&-\frac{J}{t}\Theta_j^2
\end{array}
\right),
\end{equation}
where $\Theta_j=\cos(2\pi\alpha j+\theta)$. In order to obtain an analytical expression for the Lyapunov exponent~\cite{Avila2015, Avila2017LY, SJitomirskaya2017}, we employ Avila's global theory of one-frequency analytical $SL(2,\mathbb{C})$ cocycle~\cite{{Avila2015}} to the elements within the matrix. By letting $\theta\to\theta+i\epsilon$ and under the case of large $\epsilon$ limit, we obtain
\begin{equation}\label{E11}
T_{j}(\theta)=
\frac{e^{-i2\pi\alpha j}e^{\vert\epsilon\vert}}{4\Theta_j}
\left(
\begin{array}{cc}
\frac{E^2}{Jt}&-\frac{E}{t}\\
\frac{E}{t}&-\frac{J}{t}
\end{array}
\right)+\mathcal{O}(1).
\end{equation}
Then, one can get the Lyapunov exponent
\begin{equation}
\begin{split}
\gamma(E)&=\lim\limits_{L\to\infty}\frac{1}{L}\ln\Arrowvert\prod_{j=1}^{L}T_{j}(\theta)\Arrowvert\\
&=\lim\limits_{L\to\infty}\frac{1}{L}\ln\Arrowvert\prod_{j=1}^{L}\frac{1}{\Theta_{j}}\Arrowvert+\ln\vert\frac{\frac{E^2}{Jt}-\frac{J}{t}}{4}\vert+\epsilon+\mathcal{O}(1).
\end{split}
\end{equation}
By applying ergodic theory~\cite{SLonghi2019, SJitomirskaya2017, TLiu2022, YCZhang2022, ZBWang2022}, the first term of the equation can be replaced with an integral, i.e., 
\begin{equation}
\begin{split}
\lim\limits_{L\to\infty}\frac{1}{L}\ln\Arrowvert\prod_{j=1}^{L}\frac{1}{\Theta_{j}}\Arrowvert=\frac{1}{2\pi}\int_0^{2\pi}\ln\vert\sec(\theta)\vert d\theta=\ln2.
\end{split}
\end{equation}
It is worth noting that the meromorphic function inside the integral is precisely the source of the multifractality~\cite{SJitomirskaya2017, TLiu2022, YCZhang2022, ZBWang2022}. Thus, we obtain
\begin{equation}
\gamma(E)=\ln\vert\frac{\frac{E^2}{Jt}-\frac{J}{t}}{2}\vert+\epsilon+\mathcal{O}(1).
\end{equation}
According to Avila's global theory, $\gamma(E)$ is a convex piecewise linear function with respect to $\epsilon$~\cite{Avila2015}. In large
$\epsilon$ limit, the slope of $\gamma$ is always one, which means that $\mathcal{O}(1)$ term can be ignored. Since Lyapunov exponent's expression is an even function, the energy $E$ belongs to the spectrum of Hamiltonian~\eqref{E1}, then we have
\begin{equation}
\begin{split}
\gamma(E)=\max\{\ln\vert\frac{\frac{E^2}{Jt}-\frac{J}{t}}{2}\vert+\vert\epsilon\vert,~0\}.
\end{split}
\end{equation}
Since $\epsilon$ affects Lyapunov exponents only quantitatively (increasing the overall value), not qualitatively, it is safe to drop the $\epsilon$ term from the final expression~\cite{{Avila2015}}. Finally, we obtain
\begin{equation}\label{E17}
\begin{split}
\gamma(E)=\max\{\ln\vert\frac{\frac{E^2}{Jt}-\frac{J}{t}}{2}\vert,~0\}.
\end{split}
\end{equation}
From the above expression, one can find that when the input value in natural logarithm function is greater than one, the Lyapunov exponent is larger than zero, which means $\ln|1|$ is the critical point~\cite{{Avila2015}}. Based on this, one can obtain the analytical expression of the mobility edge, i.e., 
\begin{equation}\label{E18}
\vert E^2-J^2\pm2Jt\vert=0.
\end{equation}
The analytical results agree with numerical ones well~(see black dashed line in Fig.~\ref{fig-2}(a)).

To further elucidate the theoretical origin of multifractality, we can perform an integral over the lattice points within a unit cell~\cite{HTHu2025, JMZhang2025}. According to Eq.~\eqref{E2}, we obtain an effective eigenequation
\begin{equation}
E^2\psi_j=(t_j^2+J^2)\psi_j+Jt_j\psi_{j-1}+Jt_{j+1}\psi_{j+1}.
\end{equation}
The corresponding effective Hamiltonian yields
\begin{equation}
H_{\text{eff}}=\sum_{j=1}^{L/2-1}\mathcal{J}_{\text{eff},j}(c_{j}^{\dagger}c_{j+1}+H.c.)+\sum_{j=1}^{L/2}\mathcal{V}_{\text{eff},j}c_{j}^{\dagger}c_{j}
\end{equation}
with $c_{j}^{\dagger}$ ($c_{j}$) denoting creation (annihilation) operators, while the effective hopping $\mathcal{J}_{\text{eff},j}$, effective potential $\mathcal{V}_{\text{eff},j}$, and effective energy $E_{\text{eff}}$ are given by
\begin{equation}
\begin{matrix}
\mathcal{J}_{\text{eff},j}=Jt_j,\\
\\
\mathcal{V}_{\text{eff},j}=t_j^2+J^2,\\
\\
E_{\text{eff}}=E^2.
\end{matrix}
\end{equation}
 Note that, an unbounded structure emerges here in the effective potential. According to the Simon-Spencer theorem, since the potential $\mathcal{V}_{\text{eff}}$ is unbounded at infinity, the AC part of the spectrum of $H_{\text{eff}}$ is empty, i.e., the spectrum only comprises PP and/or SC parts, which correspond to the localized phase and the multifractal phase, respectively~\cite{BSimon1989, SLonghi2023, Avila2017AC}. If the system possesses a mobility edge, i.e., there exists a region where the Lyapunov exponent vanishes (see Eq.~\eqref{E17}, ~\eqref{E18}), then this region can only be a multifractal phase.

This explains the origin of the system's multifractal phase: although the original model does not explicitly contain unbounded potentials, the unbounded structure nevertheless can equivalently induce unbounded potentials. This hidden unbounded potentials precludes the existence of AC spectrum, thereby giving rise to multifractal phase. Such a concealed mechanism further extends our understanding of the conditions that can induce multifractal phases.

\section{Mechanism of Demultifractalization}\label{Sec4}

Recent studies have indicated that quasiperiodic hopping with specific structures can counteract the disruption caused by unbounded potentials to the AC spectrum, enabling the coexistence of unbounded potentials and AC spectra, and thereby achieving protection of extended states~\cite{JMZhang2025}. Here, we deliberately adopt a converse approach: based on the conclusion from the previous section, we ask whether introducing an additional unbounded potential can cancel out the effects of this hidden unbounded potential, thus realizing another form of extended protection or demultifractalization. The answer is affirmative.

\begin{figure}[htbp]
\centering
\includegraphics[width=8.5cm]{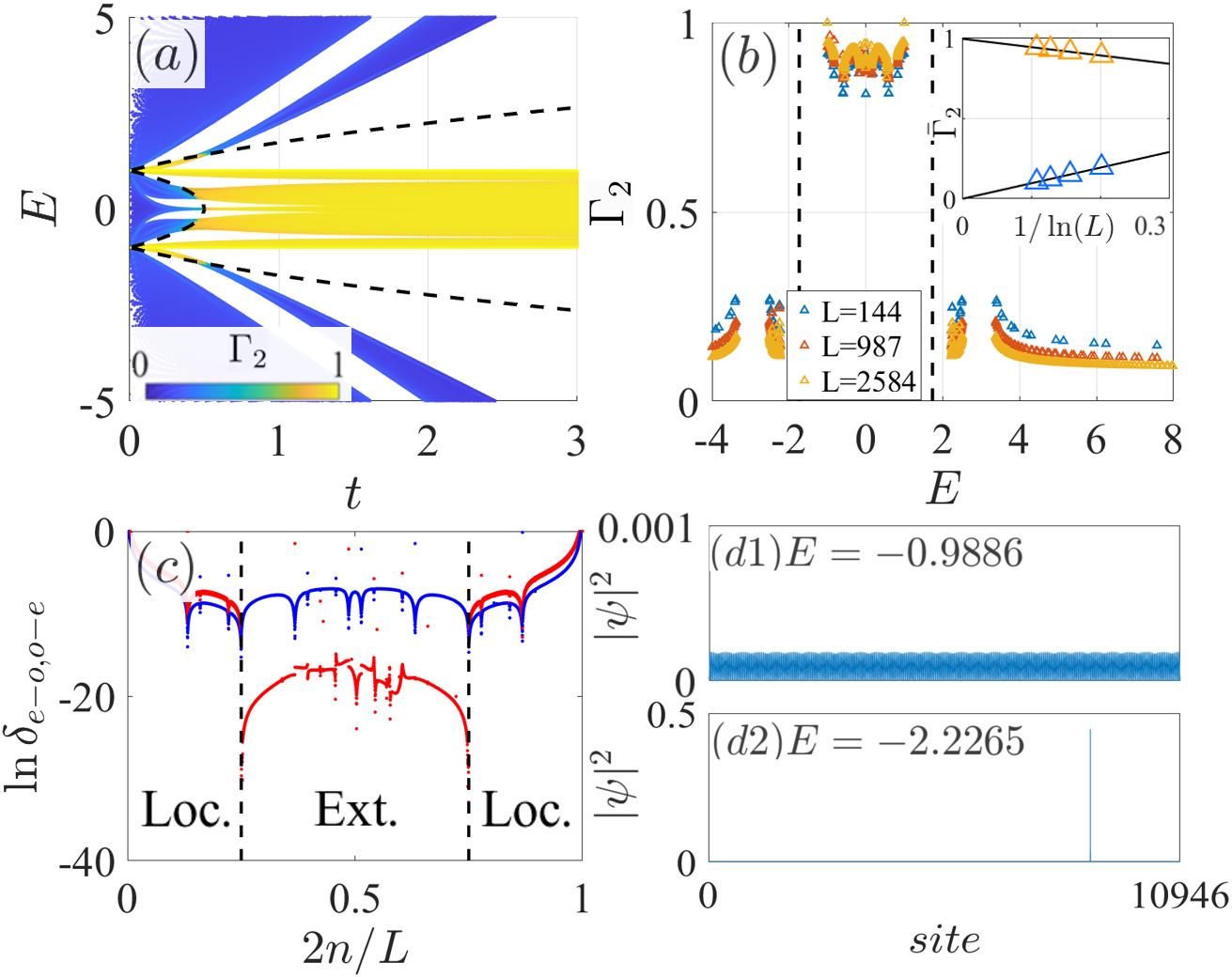}
\caption{(a) Phase diagrams of fractal dimensions in $t$-$E$ plane with $\lambda=t$. It can be seen that there exist two regions (extended or localized) where $\Gamma_2$ exhibits distinctly different behaviors, and these two regions are precisely separated by the mobility edge (black dashed line) described by Eq.~\eqref{E24}. (b) The $t=1$ cross profile of (a) for different system size. The fractal dimension of the localized region converges to 0 with increasing system size, while the extended region converges to 1. The inset exhibits the scaling behavior of extended (brown triangles) and localized (blue triangles) regions by extrapolation fitting of the mean fractal dimension of $L=$ 144, 610, 2584 and 10946. (c) The level spacings $\ln\delta^{e-o}$ (blue) and $\ln\delta^{o-e}$ (red) under the condition of $\lambda=t=1$.  The gap-opening signature characteristic of the extended phase and distinct gap characteristic of the localized phase are observed. (d) Density distribution of eigenstate in multifractal (d1) and localized (d2) region of (a). Throughout, we set $J=1$ as the energy unit. $L=2584$ for (a), while $L=10946$ for (c),(d1) and (d2).}
\label{fig-3}
\end{figure}

We now focus on the case where $\lambda=t$ (Fig.~\ref{fig-3}). As demonstrated in Fig.~\ref{fig-3}(a),  since $\Gamma_2$ takes values only close to 0 or 1, the figure appears to indicate a direct transition between localized and extended phases, separated by a mobility edge (black dashed line) described by Eq.~\eqref{E24}, with no evidence of multifractal region~\cite{HYao2019, XDeng2019, YWang2020a, YWang2022b, SZLi2024c, AJagannathan2021}. To validate the robustness of these findings across varying system sizes, we conducted a finite-size analysis~\cite{YWang2022b}, which verifies the emergence of extended states and the absence of multifractality in the case where $\lambda=t$ (Fig.~\ref{fig-3}(b)). Furthermore, the level spacing statistics reveal a clear dichotomy between localized and extended phases~\cite{APadhan2022, YZhang2022, MSarkar2021, XDeng2019, RQi2023}. Notably, the transition between even-odd and odd-even level spacing gaps is distinctly binary, either a pronounced gap exists or it is entirely absent (Fig.~\ref{fig-3}(c)). Crucially, the scattered distribution characteristic of multifractal states is not observed here, reinforcing the conclusion that such states are absent in this system. Additionally, the level spacing results confirm that the effect of the hidden unbounded potential is canceled out by the externally introduced competing potential, thereby achieving an alternative form of extended‑state protection. Following a similar approach as in the preceding text, the spatial profiles of wave functions in distinct phases are depicted in Fig.~\ref{fig-3}(d)-(e), further illustrating the sharp contrast between localized and extended states.

we can employ an analytical approach analogous to the one used earlier to further confirm the emergence of extended states. In this case, Eq.~\eqref{E11} can be rewritten as 
\begin{equation}
T_{j}(\theta)=
\left(
\begin{array}{cc}
\frac{E^2}{Jt}\Theta_j-\frac{2E\lambda}{J}&-\frac{E}{t}\Theta_j+\frac{\lambda}{t}\\
\frac{E}{t}\Theta_j-\frac{\lambda}{t}&-\frac{J}{t}\Theta_j
\end{array}
\right).
\end{equation}
Using the previously employed analytical approach~\cite{Avila2015, SLonghi2019, SJitomirskaya2017, TLiu2022, YCZhang2022, ZBWang2022}, we can obtain the corresponding Lyapunov exponent
\begin{equation}
\begin{split}
\gamma(E)=\max\{\ln\vert\frac{\frac{E^2}{Jt}-\frac{J}{t}}{2}\vert,~0\}.
\end{split}
\end{equation}
Then, we arrive at the expression for the mobility edge
\begin{equation}\label{E24}
\vert E^2-J^2\pm2Jt\vert=0.
\end{equation}
It is noted that this expression shares exactly the same algebraic form as Eq.~\eqref{E18}. However, since the integration over a meromorphic function no longer appears in the derivation~\cite{JMZhang2025, BSimon1989, SJitomirskaya2017, TLiu2022, YCZhang2022, ZBWang2022}, this mobility edge now separates localized states from extended states (rather than multifractal ones). This result also agrees well with the corresponding numerical verification, meeting our expectations.

\section{Emergent of Re-Entrant Multifractality}\label{Sec5}

\begin{figure}[htbp]
\centering
\includegraphics[width=8.5cm]{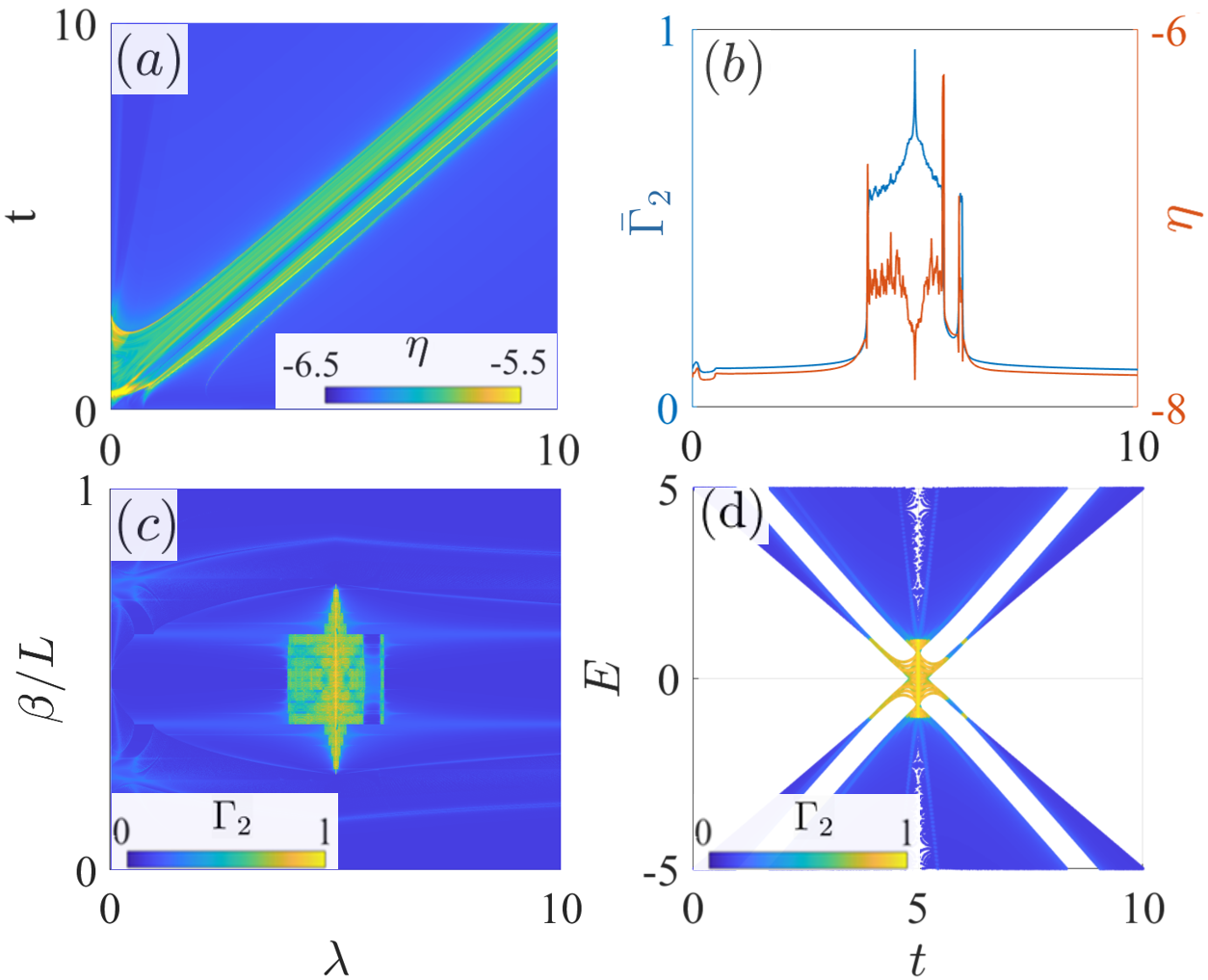}
\caption{(a) Phase diagrams of $\eta$ (defined in Eq.~\eqref{E7}) in $\lambda$-$t$ plane. The value of $\eta$ in the multifractal region is significantly larger than that in the extended or localized regions, which can serve as an indicator of the occurrence of non-monotonic phase transitions. (b) The dependence of $\overline\Gamma_2(\beta/L\in\mathcal{R})$ (defined in Eq.~\eqref{E5} and ~\eqref{E6}) and $\eta$ on the parameter $\lambda$. The two indicators collectively point to the existence of non-monotonic phase transitions. (c) and (d) Phase diagrams of fractal dimension $\Gamma_2$ in $\lambda$-$\beta/L$ plane and $\lambda$-$E$ plane, reveal a complex phase transition path: localized $\to$ multifractal $\to$ extended $\to$ multifractal $\to$ localized $\to$ multifractal $\to$ localized. We set the system size in (a) as $L=610$, while the sizes for (b), (c) and (d) are all $L=2584$.}
\label{fig-4}
\end{figure}

Based on the discussion in the preceding sections, owing to the hidden competing relationship between the quasiperiodic hopping and the quasiperiodic potential, the model described by Eq.~\eqref{E1} can be formulated to realize re‑entrant multifractalization. The corresponding results are presented in Fig.~\ref{fig-4}.

As shown in Fig.~\ref{fig-4}(a), the distribution of $\eta$ in the $\lambda-t$ plane  denotes the multifractal phase region as was discussed in Section~\ref{Sec2}~\cite{APadhan2022, XLi2017, SRoy2021, XLi2020} and reveals the competing relationship between the hopping and potential terms. It is worth noting that although $\eta$ on the diagonal $t=\lambda$ is also relatively small, it actually marks the extended phase identified in Section~\ref{Sec4}, while other regions with small $\eta$ correspond to the localized phase. Taking into account the multifractal phases in regions of larger $\eta$ near the diagonal, these results indicate that the system becomes more extended as $\lambda$ approaches $t$. In other words: when these two parameters differ significantly, the system exhibits localized behavior, as is typical for localization induced by quasiperiodic disorder. When $\lambda$ and $t$ are close to each other, the system displays multifractal phase since the system contains explicit or hidden unbounded potentials. Finally, when $\lambda=t$, the system becomes extended, which has been analytically addressed in the previous section.

In Fig.~\ref{fig-4}(b), we present the dependence of $\eta$ on $\lambda$ for the case $t=5$. Since $\eta$ exhibits consistent behavior for both extended and localized states, we supplement this with the corresponding average fractal dimension $\overline\Gamma_{\beta/L\in\mathcal{R}}$ as a function of $\lambda$, with $\mathcal{R}=[1-\alpha,\alpha]$. It can be observed that when $\lambda$ is significantly $>5$ or $<5$ , $\eta$ approaches negative infinity while $\overline\Gamma_{\beta/L\in\mathcal{R}}$ tends to 0, serving as a hallmark of the localized phase. However, when $\lambda\approx5$, $\eta$  is significantly larger than in other regions and $\overline\Gamma_{\beta/L\in\mathcal{R}}$ approaches a value between 0 and 1, indicating multifractal characteristics. Finally, when $\lambda=5$, $\eta$ again trends toward negative infinity while $\overline\Gamma_{\beta/L\in\mathcal{R}}$ tends to 1, providing clear evidence of the system's escape from the multifractal regime into the extended phase~\cite{APadhan2022, XLi2017, SRoy2021, XLi2020, HYao2019, XDeng2019, YWang2020a, SZLi2024c}. Furthermore, as shown in Fig.~\ref{fig-4}(c) (d), the Phase diagrams of average fractal dimension $\overline\Gamma_{\beta/L\in\mathcal{R}}$ at $t=5$ exhibits behavior that supports the above conclusion, demonstrating the occurrence of a re‑entrant multifractal phase. Notably, these findings also indicate the emergence of REL in the system, further enriching the results presented by this model.

To further substantiate our conclusions, we can analyze the dynamical properties of the system~\cite{ZHXu2020, MHopjan2025, SAbe1987, JXZhong2001, TMatsubara2025, GADominguez-Castro2019, ZJZhang2012, JLDong2024, AJagannathan2021}. In contrast to the previous approach, we now initialize a wave packet $\vert\psi(\tau=0)\rangle=\delta_{j,j_0}$ with $j_0=L/2$ denoting the center of the system. Using the Hamiltonian described by Eq.~\eqref{E1} and ~\eqref{E2}, the evolution equation of the system is given by $\vert\psi(\tau)\rangle=e^{-iH\tau}\vert\psi(0)\rangle$. Then, we set $t=5$ to calculate the evolution of $\vert\psi(\tau)\vert^2$ with parameter $\lambda=4.5, 5, 5.5, 5.8, 6, 6.5$, respectively. To smooth the data, it is necessary to take an average over the phase shifts $\theta$, where $\theta$ is the same as that in Eq.~\eqref{E2} and they are no longer set to zero. The phase shift considered here is $\theta_n=2\pi n/N$, where $N$ is the total number of calculations, and $n=1,2,3,\dots,N$~\cite{TMatsubara2025, ZHXu2020}. The final wave packet will be averaged over $N$ calculations to smooth the output.

\begin{figure}[htbp]
\centering
\includegraphics[width=8.5cm]{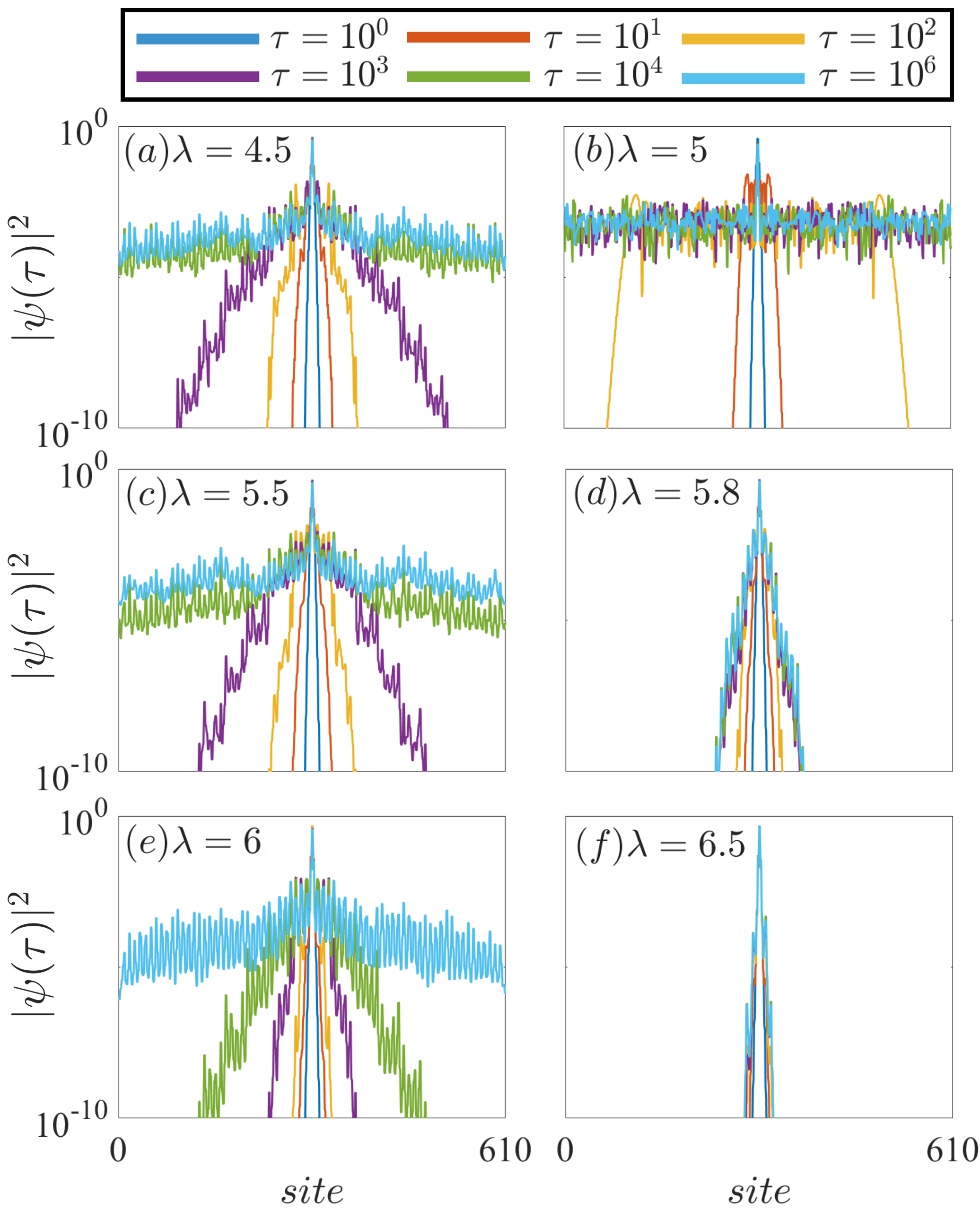}
\caption{(a)-(f) The probability distribution $\vert\psi(\tau)\vert^2$ for $t=5$ with different values of $\lambda$ after different evolution times $\tau=10^0$~(blue), $\tau=10^1$~(orange), $\tau=10^2$~(yellow), $\tau=10^3$~(purple), $\tau=10^4$~(green) and $\tau=10^6$~(light blue). For all plots, the system size $L=610$ and we choose $N=100$.}
\label{fig-5}
\end{figure}

The corresponding results are presented in Fig.~\ref{fig-5}. It is worth noting that due to the existence of a mobility edge, the system does not exhibit pure extended or multifractal phases, but rather mixed phases consisting of these states coexisting with localized phases. This means that for both extended‑localized and multifractal‑localized mixed phases, wave packets display a partially localized, partially extended character, similar to that observed in a pure multifractal phase. Nevertheless, we can still distinguish between the extended‑localized and the multifractal‑localized mixed phase by examining differences in the behavior of the extended portion of the wave packet.

Let us focus on Fig.~\ref{fig-5} (a), (b) and (c), corresponding to $\lambda=4.5$, $\lambda=5$ and $\lambda=5.5$, respectively. Although both wave packets in (a) and (b) exhibit partially localized features, the extended portion in (a) clearly shows a weaker diffusive tendency, which differs markedly from the behavior observed in (b). This indicates that (a) and (b) belong to distinct phases. Meanwhile, the behavior in (c) is essentially identical to that in (a), demonstrating that (c) and (a) reside in the same phase, i.e., the multifractal‑localized mixed phase. The results from (a)–(c) thus confirm the conclusion drawn earlier(see Fig.~\ref{fig-4}): as the quasiperiodic disorder strength increases monotonically, some eigenstates of the system escape from the multifractal regime into an extended phase, and later re‑enter the multifractal phase—a clear evidence of REM. Moreover, the wave functions in Fig.~\ref{fig-5} (d) and (f) exhibit fully exponentially localized behavior, which stands in sharp contrast to that in (e) and provides an additional clear signature of the occurrence of REL..

Based on the analytical and numerical results presented above, the emergence of REM can be understood as follows: both the explicit and hidden unbounded potentials in this system can independently induce multifractal states. However, when they coexist, a competition arises between them, leading to delocalization of the system. When their strengths are equal, the competition can even drive demultifractalization, giving rise to both REM and REL. This mechanism offers a novel perspective for further understanding Anderson localization and re‑entrant phenomena.

\section{CONCLUSION}\label{Sec6}
In summary, we study a generalized Su-Schrieffer-Heeger model with both unbounded quasiperiodical hoppings and potentials. The results demonstrate that, counterintuitively, multifractal critical phases can emerge even when only unbounded quasiperiodic hoppings are present in the system. Further theoretical analysis reveals that the hopping term in the model can be equivalent to an unbounded potential in effective eigenequation. In other words, a hidden unbounded potential is the underlying mechanism behind the emergence of critical phases. Furthermore, we introduce a conventional unbounded quasiperiodic potential to investigate the competitive interplay between it and the hidden unbounded quasiperiodic potential. The results reveal that the additionally introduced unbounded quasiperiodic potential indeed cancels the effect of the hidden unbounded potential, thereby restoring the extended phase as well as the AC spectrum. Thus, the competition between the hidden unbounded quasiperiodic potential and the traditional unbounded quasiperiodic potential can be exploited to realize reentry of the critical phase. The discoveries in this paper enrich the scope of observable phenomena in quasiperiodic systems, deepening the understanding of Anderson localization and re-entrant phenomena.

\section{Acknowledgements.}
We thank Ji-Long Dong and Shan-Zhong Li for their insightful suggestions. This work was supported by the National Key Research and Development Program of China (Grant No.2022YFA1405300), and Guangdong Provincial Quantum Science Strategic Initiative(Grant No. GDZX2304002). Yun-Yan Chen and Jia-Ming Zhang contribute equally to this work.

\appendix
\section{Definitions of observables}\label{AppA}

In order to provide a clearer introduction to the observables introduced in the main text and their underlying physical meanings, we will present a detailed and pedagogical description of the numerical methods and the definitions of these observables.

As a key quantity reflecting localization, the fractal dimension is often used to characterize the system's localization properties, and can be obtained from the scaling of the $q$ weight of the wave functions, defined by~\cite{AJagannathan2021}
\begin{equation}\label{A1}
\xi_q(\psi, \mathcal{G})=\frac{\sum_{j\in\mathcal{G}}\vert\psi_{j}\vert^{2q}}{(\sum_{j\in\mathcal{G}}\vert\psi_{j}\vert^2)^q},
\end{equation}
where $\psi_j$ denotes the projection of the wavefunction $\vert\psi\rangle$ onto the Wannier basis $\vert j\rangle$, and the sums run over all sites in a given region $\mathcal{G}$, where is generally taken to be the entire space in this work. The $q$ weight is a measure of the fraction of the presence probability
contained inside region $\mathcal{G}$, and when $q=2$, $\xi_2$ is the well-known inverse participation ratio (IPR)~\cite{AJagannathan2021}. It is not difficult to see from the mathematical form that the IPR gives the inverse of the number of lattice sites occupied by the wavefunction, which makes it a powerful tool for distinguishing wavefunctions in different phases~\cite{HYao2019, XDeng2019, YWang2020a, YWang2022b, SZLi2024c}.

The fractal dimension is defined via the scaling behavior of the $q$ weight of the wavefunction, and according to Eq.~\eqref{A1}, one can get~\cite{ XDeng2019, SZLi2024c}
\begin{equation}\label{A2}
\xi_q(\psi)\propto L^{-\Gamma_q(\psi)(q-1)}.
\end{equation}
The parameter $\Gamma_q$ is known as the $q$-th fractal dimension, denoting the power-law relationship between the 
$q$ weight of the wavefunction and the size $L$ of the region $\mathcal{G}$, and can be derived from Eq.~\eqref{A2} as
\begin{equation}\label{A3}
\Gamma_q(\psi)=-\lim_{L\rightarrow\infty}\frac{-1}{1-q}\frac{\ln\xi_q(\psi)}{\ln L}.
\end{equation}
This quantity serves as a numerical tool to distinguish the three types of eigenstates. For an extended (localized) wavefunction, one has $\Gamma_q=1(0)$ for all values of $q$, since extended states are spread uniformly over the whole space whereas localized states are frozen on a single site. For multifractal states, $\Gamma_q$ lies between 0 and 1 and varies with $q$~\cite{AJagannathan2021}. The thermodynamic limit ($L\to\infty$) results can be obtained by conducting a extrapolation fitting~\cite{YWang2022b}. In the main text, we primarily focus on the case $q=2$.

The spectrum and eigenstates of the Hamiltonian are solved by exact diagonalization, details can be found in~\cite{GADominguez-Castro2019}. Given that the system may host various eigenstates, we order their energies in ascending sequence and index them by 
$\beta$. The fractal dimension of the $\beta$-th eigenstate is then denoted by
\begin{equation}
\Gamma_q(\beta)=-\lim_{L\rightarrow\infty}\frac{-1}{1-q}\frac{\ln\xi_q(\psi(\beta))}{\ln L}.
\end{equation}
Furthermore, to compute the average fractal dimension in mixed phases, one may define $N[R]$ as the number of indices lying in a region $R$ of the $\beta$ domain. The average fractal dimension for a pure phase, or for a specific region within a mixed phase, is then given by
\begin{equation}
\overline\Gamma_q=\frac{1}{N[R]}\sum_{\beta\in R}\Gamma_q(\beta).
\end{equation}
For the purpose of identifying a specific component within mixed phases, we denote by $R=$Loc., Ext., or MF. to label the regions occupied by extended, localized, or multifractal states. Thus, the fractal dimension can be employed to characterize the localization properties of a region.

Having defined the fractal dimension above, we now turn our attention to another quantity, the normalized participation ratio (NPR) defined as~\cite{APadhan2022, XLi2017, SRoy2021, XLi2020}
\begin{equation}\label{A6}
\zeta_2(\beta)=\frac{1}{L\xi_2(\beta)},
\end{equation}
where $\xi_2$ (IPR) has already been defined in Eq.~\eqref{A1}. 

From Eq~\eqref{A1} and ~\eqref{A6}, it is evident that the IPR and NPR behave in opposite manners when characterizing the localization properties of wavefunctions. In the thermodynamic limit, for extended states, $\xi_2\sim L^{-1}\to0$ and $\zeta_2\sim L^0\to1$, whereas for localized states, $\xi_2\sim L^{0}\to1$ and $\zeta_2\sim L^{-1}\to0$. It should be noted that multifractal states are described by Eq.~\eqref{A2}, i.e., $\xi_2\sim L^{-\Gamma_2}$ and $\zeta_2\sim L^{\Gamma_2-1}$. Based on this, one may define an observable as~\cite{APadhan2022, XLi2017, SRoy2021, XLi2020}
\begin{equation}\label{A7}
\eta=\ln(\frac{1}{N[\beta/L\in\mathcal{R}]^2}\sum_{\beta/L\in\mathcal{R}}\xi_2(\beta)\times\sum_{\beta/L\in\mathcal{R}}\zeta_2(\beta)),
\end{equation}
where $\mathcal{R}$ represents the range from which eigenstates are selected. In the main text, the exact range of 
$\mathcal{R}$ is given by Eq.~\eqref{A8}. Consistent with the preceding text, $N[\beta/L\in\mathcal{R}]$ denotes the total number of eigenstates in region $\beta/L\in\mathcal{R}$.

In the thermodynamic limit, $\xi_2\times\zeta_2\sim L^{-1}\to0$ holds for extended, localized, and multifractal states alike. However, the quantity we evaluate is in fact the product of the averaged $\xi_2$ and averaged $\zeta_2$ over the region $\mathcal{R}$. If all states within $\mathcal{R}$ share the same fractal dimension—say, 
$\mathcal{R}$ lies in a pure extended or pure localized region—then one straightforwardly obtains $\eta\sim\ln L^{-1}$. In contrast, if $\mathcal{R}$ hosts both extended and localized states, this may give rise to the product of the $\zeta_2$ contributions from the extended part and the $\xi_2$ contributions from the localized part, leading to a significant number of instances occur where $\xi_2\times\zeta_2\sim L^{0}\times L^{0}=1$, and consequently $\eta$ in Eq.~\eqref{A7} approaches a finite value~\cite{APadhan2022, XLi2017, SRoy2021, XLi2020}.

In fact, in many works, this observable is commonly used to distinguish pure phases from mixed phases~\cite{APadhan2022, XLi2017, SRoy2021, XLi2020}. In this paper, however, we aim to extend its application to the identification of multifractal phases. Our attempt is based on the previous discussion on the fractal dimension. In the thermodynamic limit, the fractal dimension is 1 (or 0) for extended (or localized) phases. For multifractal phases, since they represent critical states lying between extended and localized ones, their fractal dimensions take values between 0 and 1. Moreover, the fractal dimensions of individual eigenstates within a multifractal phase are almost never identical to each other. This gives rise to an effect similar to that of a mixed phase, making the value of $\eta$ computed via Eq.~\eqref{A7} significantly larger in the multifractal phase than in the pure extended or localized phases, thereby enabling the distinction of multifractal phases from different phase regions. In the main-text calculations, to eliminate the effect of mixed phases on $\eta$, we define the region $\mathcal{R}$ as
\begin{equation}\label{A8}
\mathcal{R}\in[1-\alpha,\alpha],
\end{equation}
such that it always contains only one of the three components—namely, the extended phase, the localized phase, or the multifractal phase.

\begin{figure*}[htbp]
\centering
\includegraphics[width=17cm]{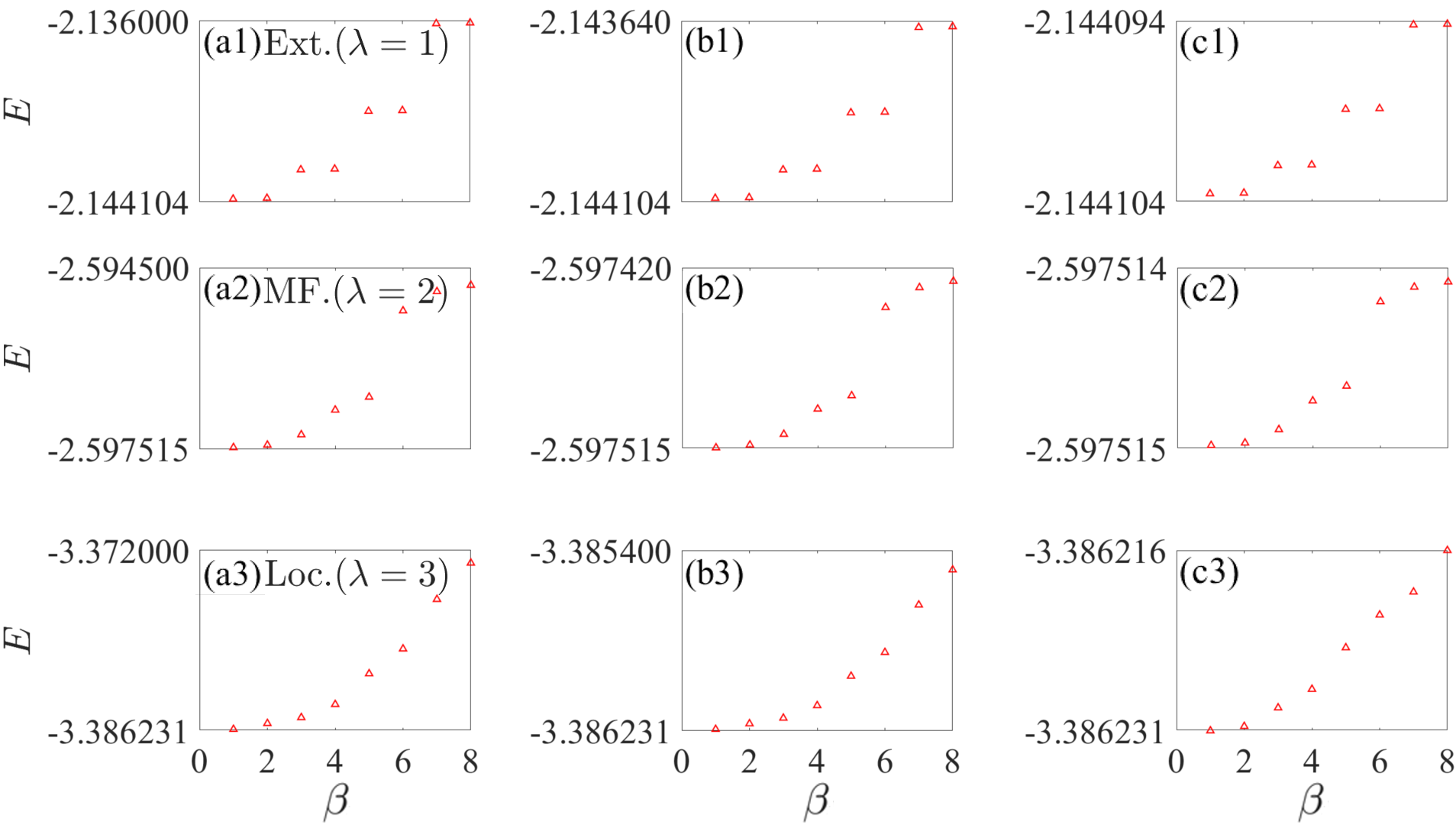}
\caption{The eight lowest eigenvalues in the spectra of standard AA model under the condition of the extended (a1-c1), multifractal (a2-c2), and localized (a3-c3) phases. From left to right columns, the system size $L=233,~987,~6765$.}
\label{fig-6}
\end{figure*}

Furthermore, the three different phases can be differentiated via the characteristics of the spectrum. Since in generalized AA models of various types, the wavefunctions associated with extended eigenstates are typically plane waves or nearly plane waves, their real and imaginary parts each represent an eigenstate with identical energy, the spectrum of extended phases exhibits doubly degeneracy~\cite{SAubry1980, APadhan2022, YZhang2022, MSarkar2021, XDeng2019, RQi2023}. Notably, this doubly degeneracy is stable only under periodic boundary conditions (PBC). In contrast, under open boundary conditions (OBC), the imaginary and real parts of plane waves are mutually exclusive, meaning that the presence of one necessarily excludes the other. We now turn to the classic Aubry-Andr\'e model to demonstrate how these spectral characteristics can be extracted. Let's consider the Hamiltonian of the AA model~\cite{SAubry1980}
\begin{equation}
H_{AA}=-J\sum_{j=1}^{L-1}(c_{j+1}^\dag c_{j}+H.c.)+\sum_{j=1}^{L}V_{AA,j}c_j^\dag c_j
\end{equation}
with
\begin{equation}
V_{AA,j}=\Lambda\cos(2\pi\alpha j+\theta)
\end{equation}
where $c_{j}(c_{j}^\dag)$ represents the fermionic annihilation (creation) operator at site $j$. Other parameters are all with the same meaning as the model Eq.~\eqref{E1} and ~\eqref{E2}. The standard AA model has precise critical points of Anderson transition, i.e., $\Lambda>,~<,~=2$ correspond to localized, extended, multifractal phases, respectively. 

Fig.~\ref{fig-6} exhibits the zoom-in plots of eigenspectrum for the extended (the top row), multifractal (the middle row), and localized phases (the bottom row) in the AA model for various system sizes. Evidently, the spectrum associated with the extended phase exhibits doubly degeneracy, whereas the doubly degeneracy is absent in the localized phase. The behavior of the multifractal phase will be discussed in more detail later. Moreover, the characteristics of the spectrum is independent of system size (see from left to right columns of Fig.~\ref{fig-6}).

In order to distinguish the three different states by the level spacing statistics, one can use the advantage of doubly degeneracy based on the above analysis. Specifically, in numerical calculations, one need to be arranged in ascending order of the level index. By subtracting the energy with odd indeces from the energy with even indices, one can obtain the all level spacing. Then, the results will form a set, namely, $\delta^{e-o}_n=E_{2n}-E_{2n-1}$. Similarly, one can obtain the other level space set of odds minus evens, i.e., $\delta^{o-e}_n=E_{2n+1}-E_{2n}$, where $n=1,2,3,..., n_{max}$ denotes the even (odd) eigenenergy in ascending order of the eigenenergy spectrum~\cite{APadhan2022, YZhang2022, MSarkar2021, XDeng2019, RQi2023}. For the case of doubly degeneracy, there will be a significant gap between $\delta^{e-o}_n$ and $\delta^{o-e}_n$. Conversely, if the double degeneracy disappears, the gap between $\delta^{e-o}_n$ and $\delta^{o-e}_n$ will disappear.

\begin{figure}[htbp]

\centering
\includegraphics[width=8.5cm]{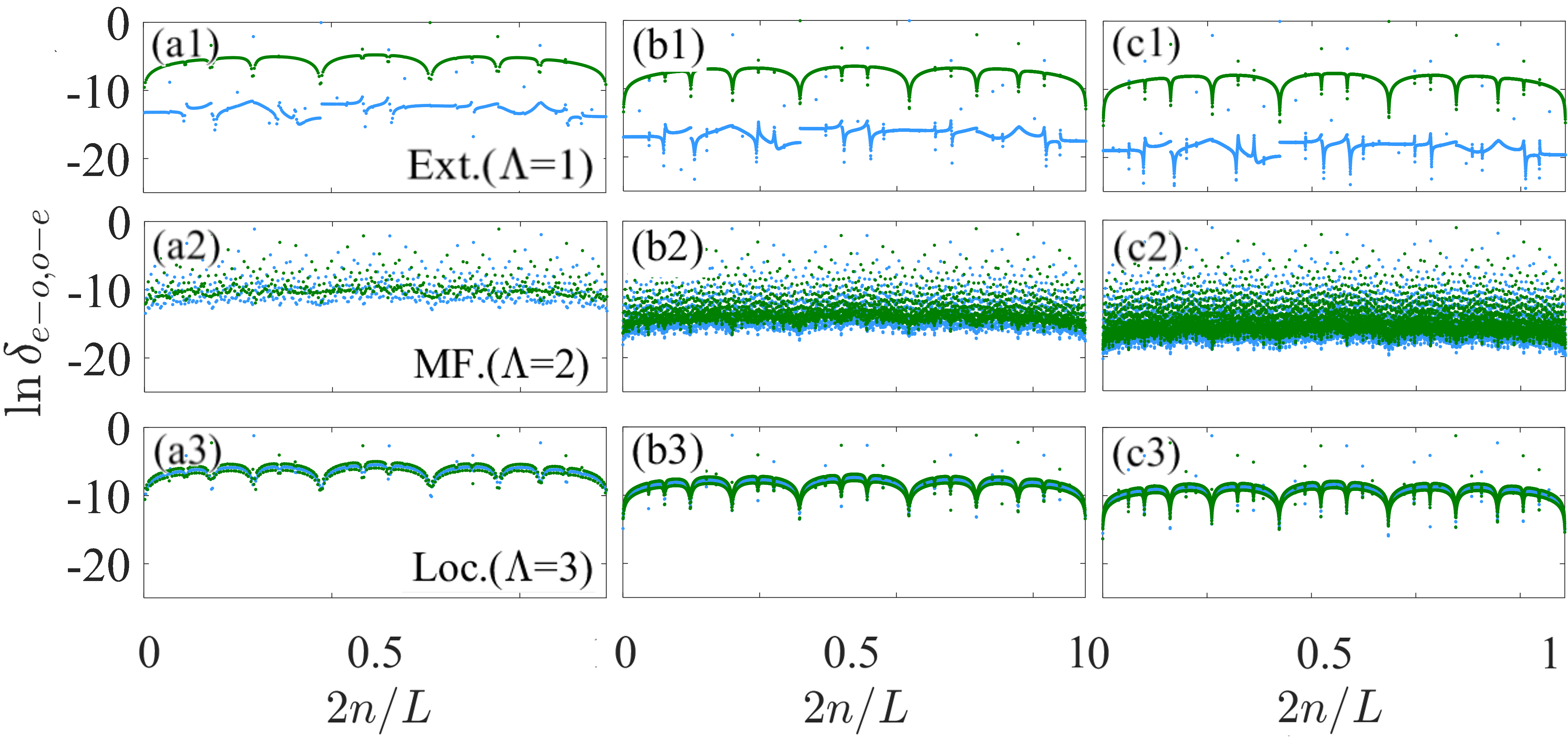}
\caption{The even-odd (light blue) and odd-even (green) level spacing of AA model under the condition of the extended (a1-c1), multifractal (a2-c2), and localized (a3-c3) phases. From left to right columns, we set $L=987,~6765,~17711$.}
\label{fig-7}
\end{figure}

However, in the thermodynamic limit, $\delta^{e-o}(\delta^{o-e})\approx0$ for each of the three phases, which makes them hard to tell apart. Yet, since the double degeneracy remains, in numerical practice one may exhibit the gap properties under large-size scenarios by taking the logarithm form~\cite{APadhan2022, YZhang2022, MSarkar2021, XDeng2019, RQi2023}. Thus, should the twofold degeneracy be present, there will be a significant gap between $\ln\delta^{e-o}_n$ and $\ln\delta^{o-e}_n$ [see Fig.~\ref{fig-7} (a1-c1)], the gap even increases as the size increases. For the case of localized phases [see Fig.~\ref{fig-7} (a3-c3)], $\ln\delta^{e-o}_n$ and $\ln\delta^{o-e}_n$ always overlap together at different system sizes. The corresponding properties of multifractal phases are in between. Specifically, for multifractal states, $\ln\delta^{e-o}_n$ and $\ln\delta^{o-e}_n$ hybridize since $\ln\delta^{e-o}_n\approx\ln\delta^{o-e}_n$ [see Fig.~\ref{fig-7} (a2-c2)]. The three markedly different behaviors serve as strong evidence to distinguish the three phases.

\section{Further numerical results}\label{AppB}

To guarantee the rigor of the analysis, a further investigation of the spectra of multifractal exponents and expansion dynamics is carried out. 

\begin{figure}[htbp]

\centering
\includegraphics[width=8.5cm]{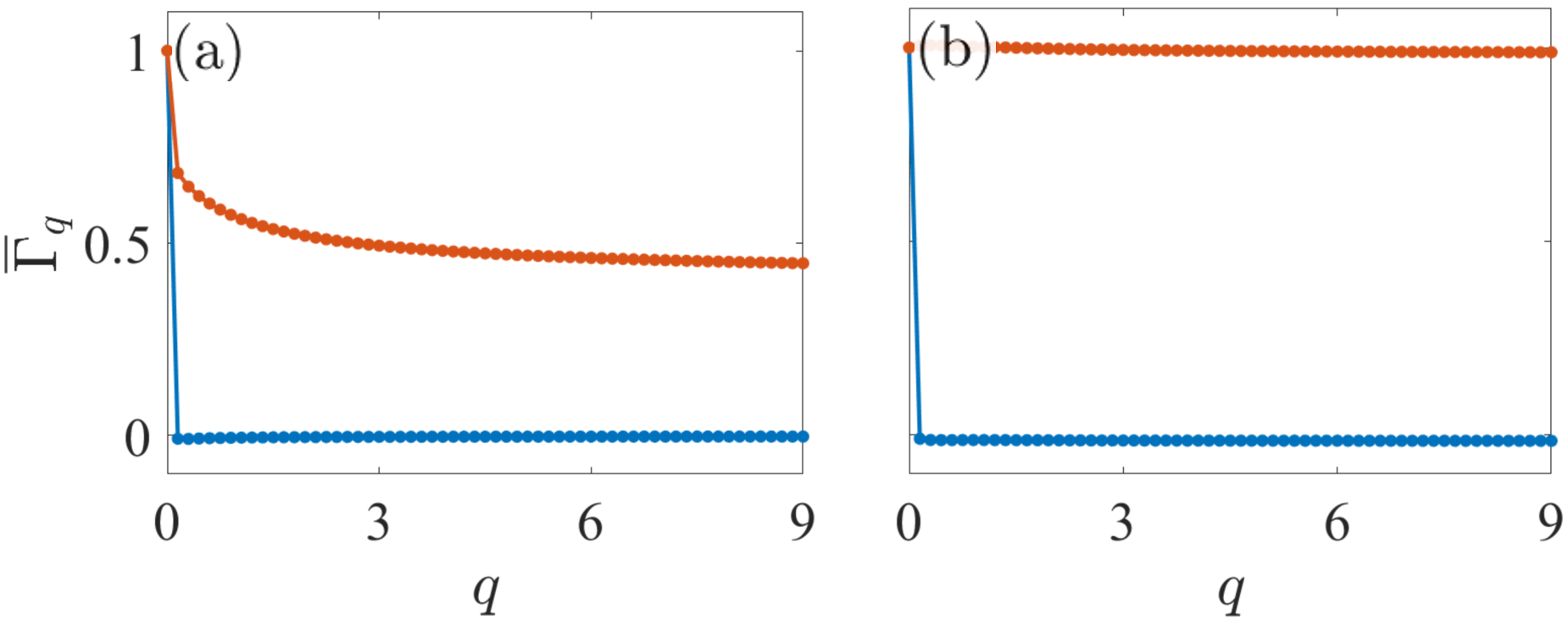}
\caption{(a) The spectra of multifractal exponents for $\lambda=0$, $t=1$, where the orange and blue dots indicate the average fractal dimensions for $R=MF.$ and $R=Loc.$, respectively. (b) is for $\lambda=1$, $t=1$, where the orange and blue dots indicate the average fractal dimensions for $R=Ext.$ and $R=Loc.$, respectively. All data points are derived from thermodynamic extrapolation (as in Fig.~\ref{fig-2} (b) and ~\ref{fig-3} (b)), with the extrapolation performed using the following system sizes $L=$ 144, 610 and 2584.}
\label{fig-8}
\end{figure}

First, taking the fractal dimension defined in Eq.~\eqref{A3}, it is known that for multifractal states $\Gamma_q$ displays a nontrivial dependence on $q$~\cite{AJagannathan2021}. For a more rigorous verification of the observation of multifractal phase, the spectra of multifractal exponents for various parameters are calculated and presented in Fig.~\ref{fig-8}.

As shown in Fig.~\ref{fig-8}, for the extended and localized phases, the average fractal dimension stays nearly constant at 1 or 0, independent of $q$. In contrast, for the multifractal phase, the average fractal dimension changes with $q$, revealing a nontrivial dependence on $q$. This provides strong evidence for the emergence of multifractal phases in the system~\cite{AJagannathan2021}.

\begin{figure}[htbp]

\centering
\includegraphics[width=8.5cm]{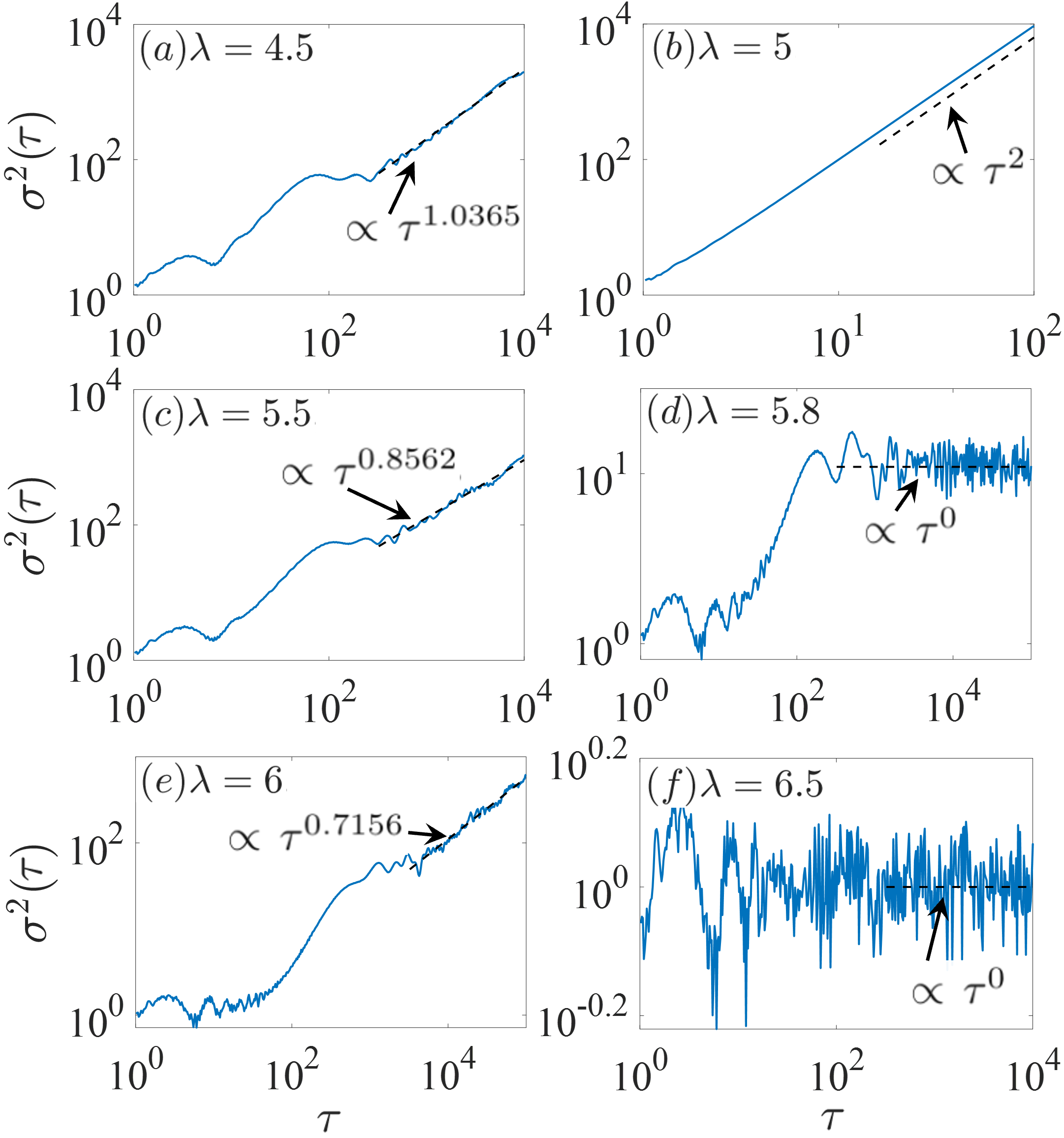}
\caption{(a)-(f) The mean-square displacement $\sigma^2(\tau)$ versus $\tau$ for different values of $\lambda$, at $t=5$. For all plots, the system size $L=610$ and we choose $N=50$. The black dashed lines in the figure indicate the fitted diffusion exponent $n$. The values obtained are $n=1.0365,~0.8562,~0.7156$ for $\lambda=4.5,~5.5,~6$, respectively, suggesting superdiffusion or subdiffusion. In contrast, for $\lambda=5$, $n\approx2$, characteristic of ballistic diffusion, and for $\lambda=5.8,~6.5$, $n\approx0$, indicating localization. It should be noted that all initial states are defined by $\vert\psi(\tau=0)\rangle=\delta_{j,j_0}$ with $j_0=L/2$.}
\label{fig-9}
\end{figure}

In addition, a further quantitative analysis of the expansion dynamics in Section~\ref{Sec5} can be carried out. The mean-square displacement $\sigma^2(\tau)$ is defined as~\cite{ZHXu2020, MHopjan2025, ZJZhang2012, JLDong2024, AJagannathan2021}
\begin{equation}
\sigma^2(\tau)=\sum_j(j-j_0)^2\vert\psi_j(\tau)\vert^2.
\end{equation}
This observable measures the diffusion speed of the wave packet, and after long-time evolution one has $\sigma^2(\tau)\propto\tau^n$, with $n$ being the diffusion exponent~\cite{ZHXu2020, MHopjan2025, ZJZhang2012, JLDong2024}. In uniform lattices corresponding to the extended phase, ballistic diffusion ($n=2$) is observed, whereas disorder leads to localization with $n=0$. Multifractal phases can exhibit either superdiffusion ($1<n<2$) or subdiffusion ($0<n<1$)~\cite{MHopjan2025, SAbe1987, JXZhong2001, ZJZhang2012, JLDong2024, ZHXu2020, MHopjan2025, AJagannathan2021}. Thus, by evaluating the relation between $\sigma^2(\tau)$ and $\tau^n$, one can quantitatively distinguish the three phases, while excluding the influence of localized contributions in mixed phases.

Fig.~\ref{fig-9} displays $\sigma^2(\tau)$ as a function of $\tau$, with the black dashed lines representing the fitted diffusion exponent $n$. The diffusion exponents extracted from Fig.~\ref{fig-9} (a), (c), and (e) are found to lie either in the superdiffusive range $1<n<2$ or in the subdiffusive range $0<n<1$. These observations suggest that, at $\lambda=4.5,~5.5$ and $6$, the system hosts a multifractal phase, lying between the extended and localized regimes. However, at $\lambda=5$ (b), the diffusion exponent is found to be $n\approx2$, signaling the presence of the extended phase. In contrast, for $\lambda=5.8$ (d) and $6.5$ (f), $n\approx0$, which is characteristic of the localized phase. These observations quantitatively confirm a non-monotonic phase transition—multifractal $\to$ extended $\to$ multifractal $\to$ localized $\to$ multifractal $\to$ localized—as $\lambda$ is increased monotonically, providing a strong complement to the discussion in Section~\ref{Sec5}.

\begin{figure}[H]
\centering
\includegraphics[width=8.5cm]{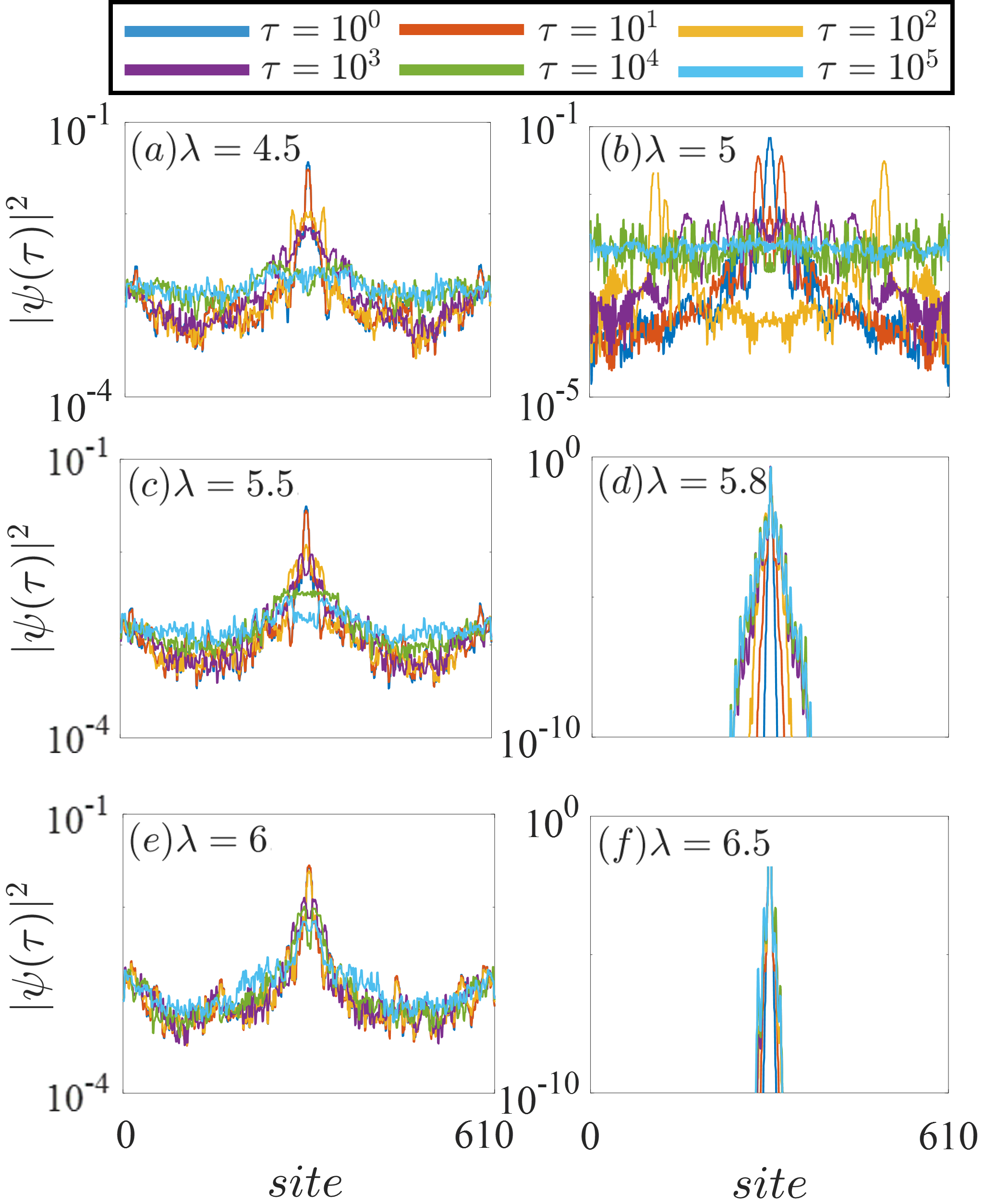}
\caption{(a)-(f) The probability distribution $\vert\psi(\tau)\vert^2$ for $t=5$ with different values of $\lambda$ after different evolution times $\tau=10^0$~(blue), $\tau=10^1$~(orange), $\tau=10^2$~(yellow), $\tau=10^3$~(purple), $\tau=10^4$~(green) and $\tau=10^5$~(light blue). For all plots, the system size $L=610$ and we choose $N=100$. In contrast to Fig.~\ref{fig-5}, for $\lambda=4.5,~5,~5.5,~6$, the initial states are replaced by those defined in Eq.~\eqref{B2} with $E_0=0$ and $r=50$, thereby enabling the extraction of multifractal or extended phases components through energy filter.}
\label{fig-10}
\end{figure}

Finally, we provide a brief supplement to the results in Fig.~\ref{fig-5}. One can set the initial state of the time evolution as linear superposition of eigenstates of the Hamiltonian, thereby obtaining an initial wave packet that is approximately localized near the center of the system with a specific energy distribution, i.e.,
\begin{equation}\label{B2}
\vert\psi(\tau=0)\rangle=\sum_{\beta=\beta(E_0)-r}^{\beta(E_0)+r}\vert\psi(\beta)\rangle,
\end{equation}
where $E_0$ represents the center of the energy window, $\beta(E_0)$ refers to the index associated with $E_0$ , and  $r$ is the window width measured in the index space of $\beta$. Replacing the initial states for $\lambda=4.5,~5,~5.5,~6$ in Fig.~\ref{fig-5} by those specified in Eq.~\eqref{B2} with $E_0=0$ and $r=50$, one can yields the time evolution of the individual components in the mixed phases, which are presented in Fig.~\ref{fig-10}.

As is clearly seen in Fig.~\ref{fig-10} (b) ($\lambda=5$), the wave packet corresponding to the extended state spreads rapidly and eventually becomes uniformly distributed over the whole space. By contrast, Fig.~\ref{fig-10} (d) ($\lambda=5.8$) and (f) ($\lambda=6.5$) exhibit localized states, where the wave packet remains confined near a localized center and decays exponentially with distance. The wavefunctions in Fig.~\ref{fig-10} (a) ($\lambda=4.5$), (c) ($\lambda=5.5$), and (e) ($\lambda=6$), however, are intermediate between the two extremes, showing neither complete localization nor full extension, which is characteristic of multifractal dynamics. These observations further corroborate the non-monotonic phase transition discussed in Section~\ref{Sec5}.

\end{document}